\documentclass[]{aa} % for a referee version
\usepackage{graphicx}
\begin{document}
   \title{An optimal extraction of spatially blended spectra}

   \subtitle{}

   \author{ R. I. Hynes }

   \institute{Department of Physics and Astronomy, University of Southampton, 
              Southampton, SO17 1BJ, UK\\
              \email{rih@astro.soton.ac.uk} 
   }

   \date{Received 9 October 2001 / Accepted 26 November 2001 }

   \abstract{
   We describe an algorithm to optimally extract individual
   spectra of blended sources from a long slit spectrum.  A semi-analytic
   model for the spatial profile is used: a Voigt profile for the
   undersampled core with a numerical correction applied to the wings.
   This is derived from an isolated source which must also be placed on the
   slit.  $\chi^2$ fitting is used to separate the blended components
   with maximum weight given to the best data.  We demonstrate the
   successful application of the algorithm to data on two X-ray binaries
   in crowded fields: \object{XTE~J2012+381} and \object{V404~Cyg}.
   }

   \maketitle
%
%________________________________________________________________

\section{Introduction}
\label{IntroSection}
A common situation in observational astronomy is to seek accurate
photometric or spectroscopic information about sources within crowded
fields, i.e.\ where the nearest neighbours are blended with the
sources of interest.  This is a particularly common situation for X-ray
binaries, most of which are faint Galactic plane objects.  A solution
to the photometric problem has long been available in various
implementations of Stetson's {\sc daophot} program (Stetson
\cite{Stetson:1987a}).  For crowded field spectroscopy, however, no
widely available solution exists.  A possible approach to the problem
is described here and has much in common with {\sc daophot}.  This
spectroscopic solution also seeks to be {\em optimal}, in the sense
that it should yield the highest signal-to-noise spectra available
from the data.  It thus derives from optimal extraction methods
developed for unblended spectra (Horne \cite{Horne:1986a}; Marsh
\cite{Marsh:1989a}).  We developed the method initially in the context
of an observation of the X-ray binary \object{XTE~J2012+381} (Hynes et
al.\ \cite{Hynes:1999a}).  It has also provided useful for another
X-ray binary, \object{V404~Cyg} (Hynes et al.\ \cite{Hynes:2002a}).
We illustrate the application of the method to both these sources in
Section \ref{ExampleSection}.

Since developing the method, another approach has been presented by
Buie \& Grundy (\cite{Buie:2000a}).  This is also an optimal
extraction, but differs in that no template source is used; instead
the spatial profile is reconstructed iteratively from the blended
sources.  These authors also use a purely numerical profile, rather than
the semi-analytic model we adopt.  Their approach was developed for
{\it HST}/NICMOS observations of a Pluto and Charon; hence these
requirements were necessary.  For ground-based observations of point
sources in crowded fields, using a template source and a semi-analytic
model profile should allow our method to be applied to poorer quality
data.  We should emphasise that the assumptions made here do only
apply to point sources; this method cannot be used for sources with
extended or resolved structure.
%
%%%%%%%%%%%%%%%%%%%%%%%%%%%%%%%%%%%%%%%%%%%%%%%%%%%%%%%%%%%%%%%%%%%%%%%%%%%%%%%
%
\section{Preliminary Image Processing}
\label{ReductionSection}
Before applying the method, the images should have been debiased and
flat-fielded using standard techniques.  It is assumed hereafter that
the image is oriented with the dispersion direction roughly horizontal
and the slit is aligned with columns.

Sky subtraction should also have been performed and to ensure correct
weighting across the profile, this should be done in a way which
yields a image of the fitted sky background.  Horne
(\cite{Horne:1986a}) discusses these issues and we essentially follow
his methods.  Our implementation fits a low-order polynomial to each
column with Horne's iterative cosmic ray rejection.  Regions
containing sources are masked out.  We then retain both the subtracted
image and an image containing the fitted sky.

It is also necessary to do wavelength calibration separately using
normal procedures.  The method employed for the spectra presented in
Section \ref{ExampleSection} was to extract and wavelength calibrate a
spectrum of the template source using normal {\sc iraf} methods, and
then assume the same wavelength--pixel correspondence for all the
objects.  This is acceptable for these data, since the spectra
considered do have the image of the slit precisely aligned with the
CCD columns, but in general a more rigorous procedure may be needed.
%
%%%%%%%%%%%%%%%%%%%%%%%%%%%%%%%%%%%%%%%%%%%%%%%%%%%%%%%%%%%%%%%%%%%%%%%%%%%%%%%
%
\section{The Spatial Profile}
\label{ProfileSection}
\subsection{The profile model}
The first step in deblending spatial profiles is to define how the
profile of a single point source should look.  Several methods have been
proposed for this for a single source.  Horne (\cite{Horne:1986a})
normalises the observed count rates for each column, then fits a low
order polynomial to each row of the data.  This has the advantage of
making no {\it a priori} assumptions about the spatial profile, but is
inappropriate for spectrum deblending because his spatial profile is
only defined in pixel steps; it cannot easily be transferred to
another source which samples the spatial profile differently.  In
principle, one could interpolate between pixels, but this is dangerous
when the spatial profile is undersampled, as is often the case (even
where the seeing is poor enough that the spatial profile would become
oversampled, it is common to bin to reduce readout noise).  Horne's
method is also only appropriate for spectra with small distortion,
i.e.\ nearly parallel to CCD rows.  Marsh (\cite{Marsh:1989a})
describes an extension of this empirical approach which works even for
very distorted spectra.  His method also produces well sampled spatial
profiles which can be transferred to another source on the slit, taking
advantage of the fact that strong distortion of the spectrum will lead
to different columns sampling the spatial profile differently.  This
could be used as the basis for a deblending algorithm.  In some cases,
however, the distortion is very slight so this method will not work
either; even by combining columns, the spatial profile is not well
sampled.  This was the case for the two datasets discussed in Section
\ref{ExampleSection}, and so led us to develop a different method.

The alternative is simpler to visualise, since it involves fitting a
analytical profile in the spatial direction rather than a polynomial
in the dispersion direction.  Analytic spatial profiles have been
applied by Urry (\cite{Urry:1988a}) to {\it IUE} data using a Gaussian
profile.  For ground based data a Gaussian model will often give a
good fit to the seeing dominated cores of spatial profiles, but will
usually underpredict the extended wings present due to instrumental
imperfections.  An effective refinement to this is to use a Voigt
profile, a convolution of a Gaussian profile with a dispersion profile
(e.g.\ Gray \cite{Gray:1992a}):
\begin{equation}
P_i(u_i,a) \propto 
\int^{\infty}_{-\infty} \frac{\exp(-u_1^2)}{(u_i-u_1)^2+a^2}{\rm d}u_1.
\end{equation}
where $P_i$ is the profile value, $u_i=x_i/w$, $x_i$ is the pixel
position, $w$ is the profile width parameter and $a$ is the Voigt
damping parameter.  This function provides the desired Gaussian cores,
but also gives an independently variable extension in the wings,
controlled by the Voigt damping parameter which ranges from 0 (a pure
Gaussian profile) to 1 (a Lorentzian profile).  It was found that a
Voigt profile provided a very good fit to the spatial profiles
considered in Section \ref{ExampleSection}.  The resulting systematic
residuals to the fit are very small; a numerical correction method
similar to that used by {\sc daophot} can be used to refine the fit
even more as described in the following subsection.
\subsection{Fitting the profile}
The first step in the fitting process is to perform an unconstrained
fit to each column of the template data, with the centre, Gaussian
width, Voigt damping parameter and profile scaling as free parameters.
We have used a downhill simplex (amoeba) algorithm to obtain the
best-fit parameters (e.g.\ Press et al.\ \cite{Press:1992a}).  No
attempt is made to reject bad pixels or cosmic rays at this stage, as
these will show up as anomalous parameter values.  Of these parameter
values, it is the profile centre, width and damping that define a {\it
normalised} profile.  All of these parameters can be expected to vary
smoothly in wavelength, so after obtaining fitted values, a low order
polynomial in wavelength is fitted to each parameter, rejecting
anomalous values that may have been affected by cosmic rays.  If the
wavelength range is small then a zero-order fit (i.e.\ a constant) may
suffice.  If the data quality is not sufficient to constrain the fits
to a single row of the image then several rows can be binned, since
the wavelength dependence is likely to be slow, and the derived
polynomial resampled back to the original resolution.  These
polynomial fits, together with the normalisation such that $\sum_i P_i
= 1$ define the spatial profile as a function of wavelength.

If the difference in brightness of the two sources is very large then it
becomes crucial that the wing of the brighter source be very well fitted
to avoid contamination of the fainter source.  This can be achieved
using a numerical correction to the (Voigt) model profile.  The error
involved in assuming a Voigt profile is usually only important in the
extreme wings of the profile: the seeing-dominated core is typically
well fitted by this model.  We can therefore define a numerical
correction factor (observed profile divided by model) at the sampled
points and interpolate this to obtain a general correction function
for any pixel sampling.  Since this involves the extreme wings of the
line, where signal-to-noise will be poor, such a correction typically
cannot be defined meaningfully as a function of wavelength; instead
only a single, averaged, wavelength independent correction can be
determined.  A symmetric profile is also assumed.  These assumptions
appear adequate for the spectra we have considered; hence the
correction is simply a function of distance from the profile centre.
Given higher quality data, an asymmetric, wavelength dependent
function could readily be used, and this generalisation is
straightforward.

Assuming that there are no changes in focus along the slit (or at
least the restricted part of it which is used) and that the image
scale (and hence the separation between stellar images) is independent
of wavelength, the only additional information required to define the
profiles of the blended sources is the separation of each from the
template source.  For spectra with minimal distortion, this is easy to
obtain by performing an initial fit to the average of several rows of
data, leaving the centres of the two blended profiles as free
parameters.  Distorted spectra would require that the distortion
(known from the fits to the template profile) be removed by resampling
the data for this initial fit only.
%
%%%%%%%%%%%%%%%%%%%%%%%%%%%%%%%%%%%%%%%%%%%%%%%%%%%%%%%%%%%%%%%%%%%%%%%%%%%%%%%
%
\section{Spectral Extraction}
\label{ExtractionSection}
Assuming that the fit to the spatial profile for the template source is
satisfactory, the centre position and normalised profile as a function
of wavelength are known, and the spectral extraction itself can be
performed.  This is simply a case of finding the scaling of each
component, although complicated by the need to reject cosmic ray
events.  Horne (\cite{Horne:1986a}) describes his method in terms of
an optimally weighted sum of counts over an aperture containing all of
the stellar light.  He notes the equivalence between this weighted sum
method and fitting a known profile to data of known variance, and it
can be shown that his optimally weighted sum yields the solution of
minimum $\chi^2$.  Using this equivalence we can generalise the method
to blended spectra, by minimising the $\chi^2$ of the fit with respect
to the two profile scalings.  This is straightforward and for
the simple problem of scaling two profiles, the solution of minimum
$\chi^2$ can be calculated analytically; see Appendix \ref{App}.  This
is done iteratively, to allow rejection of cosmic rays.  Given that
there are two components to be fitted, however, the {\em combined}
profile is unknown, and so more care is needed in cosmic ray rejection
than for a single source.
%
%%%%%%%%%%%%%%%%%%%%%%%%%%%%%%%%%%%%%%%%%%%%%%%%%%%%%%%%%%%%%%%%%%%%%%%%%%%%%%%
%
\section{Cosmic Ray Rejection}
\label{CosmicRaySection}
One of the great strengths of optimal extraction algorithms is that
the spatial profile is known.  Cosmic rays or bad pixels, even when on
top of the spectrum, will distort the spatial profile, and hence can
be recognised and rejected.  This means that instead of removing
cosmic rays from the extracted one-dimensional spectra, and rejecting
a whole pixel step in wavelength, they can be removed from the
two-dimensional spectra, retaining the information in the
uncontaminated pixels.  While this process is easy to do by eye, it
proves difficult to train a computer to recognise cosmic rays
automatically, especially when multiple profiles are being fit.  Given
a first approximation to the profile, bad values can be rejected by an
iterative sigma-clipping algorithm (Horne \cite{Horne:1986a}).  The
strategies we have used to obtain the crucial first approximation are
described here.

Beginning with the simplest case of recognising contamination of a
single profile, we rely on the fact that the scaling
of the profile can be estimated from a single pixel value; hence each
pixel gives an independent estimate of the integrated flux.  A
contaminated pixel will then give an anomalous estimate.  If this
estimate is severely wrong, as is often the case with cosmic rays,
then a straight average may be very much in error, leading to
subsequent rejection of good data.  A better method is to take the
median of the single pixel estimates as the first estimate for the
integrated flux.  This then gives a first approximation to the fitted
profile which forms the basis for subsequent refinement with iterative
sigma-clipping.

Where two independently fitted profiles are involved more care is
needed.  The only case considered here is that one source is brighter
than the other at all wavelengths in the spectrum (true for all data
the method has so far been applied to.)  For the brightest source, a
profile region is selected from the midpoint between the two sources to
an arbitrary cutoff on the other side.  Within this region, the
profile of the brighter source is expected to dominate, so the median
ratio method described above for single sources can be applied.  This
gives an approximate flux scaling for the brighter stellar profile,
which can then be subtracted.  The median ratio method is then applied
to the remaining profile of the fainter source to estimate a
flux scaling for this.  With an approximate scaling for both
profiles, iterative sigma-clipping can be used to refine the list of
rejected pixels.

Clearly if three or more blended sources are involved then the situation
is more complex, and there are more possible permutations of relative
brightness to consider.  This makes it harder to treat generally and
specific cases will need to be considered individually.
%
%%%%%%%%%%%%%%%%%%%%%%%%%%%%%%%%%%%%%%%%%%%%%%%%%%%%%%%%%%%%%%%%%%%%%%%%%%%%%%%
%
\section{Examples}
\label{ExampleSection}
\subsection{XTE~J2012+381}
\object{XTE~J2012+381} presents an especially difficult challenge for
spectroscopy.  The X-ray source was originally identified with what
appeared to be a normal F star.  It was only upon closer examination
of images of the field taken in good seeing that it became clear a
second fainter star was present (Hynes et al.\ \cite{Hynes:1999a}).
This was heavily blended with the F star, but lay closer to the
precise radio counterpart and now appears the most likely optical
counterpart to the X-ray source.  We obtained spectroscopy of these
two stars with the William Herschel Telescope (WHT) on 1998 July 20,
with the slit aligned to pass through both, as well as through a third
unblended star.  These data provided the original motivation for
development of this method.  A fuller description of the data is
provided by Hynes et al.\ (\cite{Hynes:1999a}).  We focus here on
aspects relevant to the deblending algorithm.

\begin{figure}
\resizebox{\hsize}{!}{\rotatebox{90}{\includegraphics{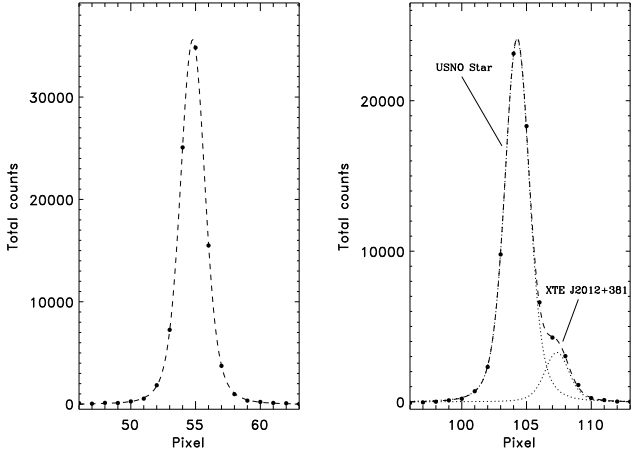}}}
\caption{Fit to spatial profiles in the red part of the WHT spectrum
of \object{XTE~J2012+381}.  Data are shown by points, the model profile is
dashed and the dotted lines indicate the De-blended components.  Both
data and fits have been summed over 20 pixels in the dispersion
direction to reduce noise and illustrate the quality of fit achieved.
On the left is the fit to the spatial profile of the template star.
On the right is the two-profile fit to the blended stars.}
\label{XTEProfileFig}
\end{figure}

\begin{figure}
\resizebox{\hsize}{!}{\includegraphics{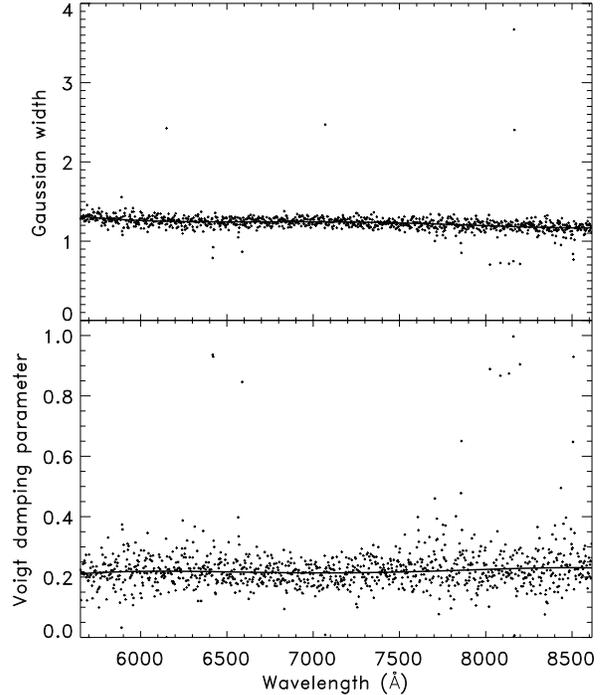}}
\caption{Wavelength dependence of the Voigt profile fitting
parameters, the Gaussian width and the Voigt damping parameter.
Points mark fits to individual pixels in the dispersion direction,
with no binning.  The solid lines are low-order polynomial fits to the
parameters.}
\label{ParameterFig}
\end{figure}

The severity of the blending can be clearly seen in Fig.\
\ref{XTEProfileFig}.  The suggested counterpart to the X-ray source is
the fainter of the two blended stars.  Clearly to successfully extract
its spectrum requires a good fit to the wing of the brighter star.
Fortunately, the signal-to-noise ratio of the two-dimensional image
was quite good, so it was possible to apply the full model described
above, i.e.\ a Voigt profile with a numerical correction to the wings.
The latter significantly improved the fit and with this no significant
discrepancy was seen between the model profile and the data.  The fits
to the Voigt profile parameters are shown in Fig.\ \ref{ParameterFig}.
Both the width and damping parameter are relatively well constrained
as a function of wavelength.

\begin{figure}
\resizebox{\hsize}{!}{\rotatebox{90}{\includegraphics{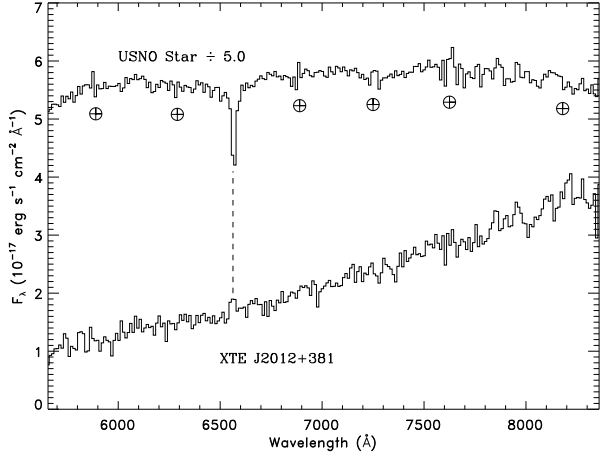}}}
\caption{WHT spectra of the F star and the faint red companion
believed to be the optical counterpart of \object{XTE~J2012+381}.  Both have
been binned $\times4$ in wavelength for clarity.  Atmospheric
absorption features (marked $\oplus$) have been corrected for in both
spectra.  The only prominent line in either spectrum is H$\alpha$:
absorption in the brighter star and possible weak emission in the
suggested counterpart to the X-ray source.}
\label{XTESpecFig}
\end{figure}

The extracted spectra are shown in Fig.\ \ref{XTESpecFig}.  There are
a shortage of spectral features in both spectra, but the extraction
does clearly distinguish the much redder spectrum of the proposed
X-ray source and the contrast between H$\alpha$ absorption in the F
star and possible weak emission in the X-ray source is dramatic.  The
reality of this emission is discussed by Hynes et al.\
(\cite{Hynes:1999a}); suffice here to note that no obvious artifacts
are present, but that these data are perhaps not the best test of the
method.  Fortunately our second target, \object{V404~Cyg}, does
provide a better test, albeit in a less heavily blended situation.

\subsection{V404~Cyg}
\begin{figure}
\resizebox{\hsize}{!}{\rotatebox{90}{\includegraphics{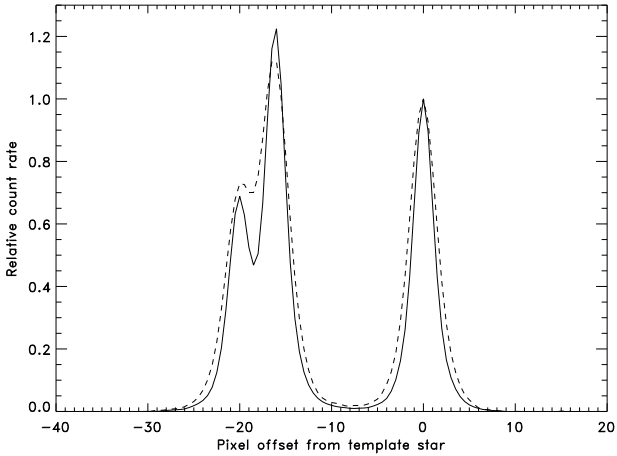}}}
\caption{Best (solid) and worst (dashed) average spatial profiles of
\object{V404~Cyg} from 1999 July 7/8.  These are the observed profiles
not models.  Columns containing cosmic rays have been excluded from
the average.  Peaks are from left to right, the blended companion
star, \object{V404~Cyg} and the profile template star.  The profiles
have been scaled such that the peak flux from the template star is
unity.}
\label{V404ProfileFig}
\end{figure}

Another X-ray binary for which this deblending method is appropriate
is \object{V404~Cyg}, as this has a fainter companion star 1.5\,arcsec
away.  There is no physical association between the two.  Observations
were obtained of \object{V404~Cyg} in quiescence using the WHT on 1999
July 6--8.  These are described in more detail in Hynes et al.\
(\cite{Hynes:2002a}).  To summarise, the goal was to perform
spectroscopy near H$\alpha$ with as high a time resolution as possible
( $\sim240$\,s) to examine spectral variability; hence
we used a low spectral resolution to minimise readout noise and a wide
slit to minimise slit losses.  Figure \ref{V404ProfileFig} shows two
examples of averaged (in wavelength) spatial cuts with respect to the
centre of the template star.  The images with the best and worst
seeing from the night of July 7/8 were chosen.  The fainter star is
clearly blended with \object{V404~Cyg} even in the best image (seeing
$\sim0.8$\,arcsec.)

The optimal deblending algorithm was applied with a simplification.
Because the goal of this project was to obtain the highest
time-resolution practical, the signal-to-noise ratio of individual
frames was not high.  Because of this, it was not possible to perform
a wavelength dependent fit and so a single average profile (and a
profile correction) were calculated.  Given the limited wavelength
coverage of these data this is an acceptable approximation.

\begin{figure}
\resizebox{\hsize}{!}{\rotatebox{90}{\includegraphics{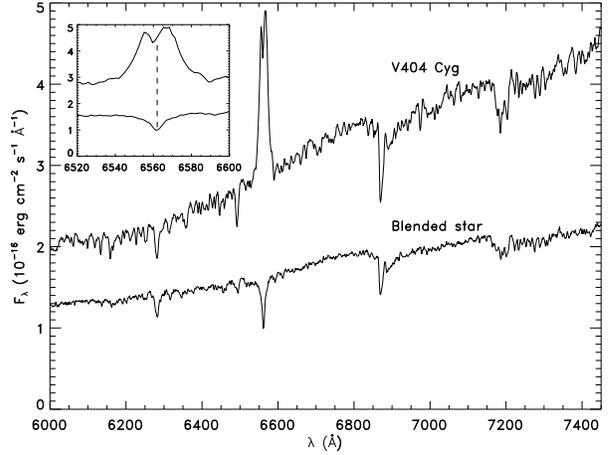}}}
\caption{Deblended spectra of \object{V404~Cyg} and the blended star.
All the spectra from 1999 July 7/8 have been averaged.  The
deblending algorithm has appears to have separated the spectra cleanly
with the H$\alpha$ emission from \object{V404~Cyg} clearly
distinguished from absorption in the blended star.  The absorption
feature in the blended star is not exactly aligned with the dip in the
line profile of \object{V404~Cyg}.}
\label{V404SpecFig}
\end{figure}

Figure \ref{V404SpecFig} shows the average extracted spectra of the
two blended stars from July 7/8.  The spectra appear to have been
separated cleanly and there is no obvious crosstalk around
H$\alpha$ between the
double-peaked disc emission in \object{V404~Cyg} and the narrower (stellar)
absorption line in the blended star.
There is also a lot of structure in the spectrum
of \object{V404~Cyg} (e.g.\ 6600--6900\,\AA) which is not present in
the spectrum of the blended star which appears largely featureless.
As \object{V404~Cyg} is brighter, the signal-to-noise ratio of its
spectrum should be higher and hence these features must be real; they
likely arise from the K0\,IV companion star which dominates in
quiescence.  The blended star, in contrast, appears to be of earlier
spectral type, probably F (Hynes et al.\ 
\cite{Hynes:2002a}).\footnote{To the author's knowledge, no spectrum
of this star has previously been isolated.}.

\begin{figure}
\resizebox{\hsize}{!}{\includegraphics{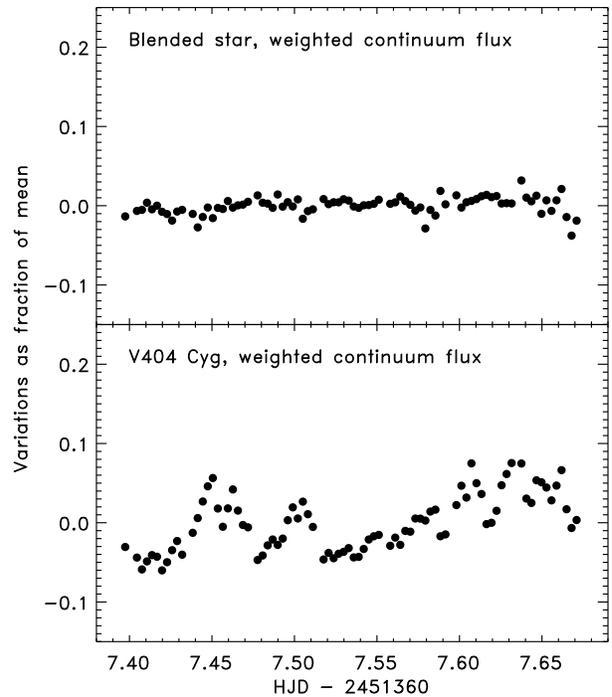}}
\caption{Continuum light curves of \object{V404~Cyg} and its blended
companion.}
\label{V404LightcurveFig}
\end{figure}

Figure \ref{V404LightcurveFig} shows continuum light curves from both
stars for the night of July 7/8.  Spectra were scaled relative to
the template star to approximately remove variable slit loss and
transparency effects.  
%Variations are seen in \object{V404~Cyg} but
%the light curve of the blended star is largely smooth.
%, though not flat
%indicating that there are calibration issues remaining to be resolved;
%simultaneous JKT photometry was obtained for this purpose, and after
%applying a correction the rise at the end of the night is removed
%(Hynes et al.\ \cite{Hynes:2002a}).  
Clearly the light curve of the
blended star shows little or no contamination from the large flare
occurring at the beginning of the night in \object{V404~Cyg},
indicating again that there is no detectable cross-talk between the
two spectra.
%
%%%%%%%%%%%%%%%%%%%%%%%%%%%%%%%%%%%%%%%%%%%%%%%%%%%%%%%%%%%%%%%%%%%%%%%%%%%%%%%
%
\section{Conclusions}
We have demonstrated an algorithm for extracting separate spectra of
sources whose spatial profiles are blended in a long slit spectrum
given an isolated template source with which to define the spatial
profile.  We implement this method using $\chi^2$ fitting with
rejection of cosmic rays, resulting in a generalisation of the optimal
extraction method to the multi-object case.  Finally we have
demonstrated the application of the method to two X-ray binaries in
crowded fields that require separation from nearby contaminating
sources.

The method as presented here has some limitations:
\begin{enumerate}
\item
A very simple method of wavelength calibration is used, i.e.\ assuming
that the slit is exactly aligned with CCD columns.  A more rigorous
treatment would be able to use a two-dimensional wavelength solution
when necessary.
\item
The assumption of a Gaussian core to the profile may not be valid if
the image size is determined by poor focus rather than seeing, or
if the exposure time is short enough ($\la1$\,s) that the
seeing does not average to a Gaussian form. 
\item
At present the method is targeted at two blended profiles.  A
generalisation to three or more sources would not only require an
extension of Appendix \ref{App}, but development of a method for
identifying cosmic rays superposed on such a complex profile.
\end{enumerate}

An implementation of the method described using IDL routines operating
on FITS images is available from the author.
%
%%%%%%%%%%%%%%%%%%%%%%%%%%%%%%%%%%%%%%%%%%%%%%%%%%%%%%%%%%%%%%%%%%%%%%%%%%%%%%%
%

\begin{acknowledgements}
The author is supported by grant F/00-180/A from the Leverhulme Trust.
The William Herschel Telescope is operated on the island of La Palma
by the Isaac Newton Group in the Spanish Observatorio del Roque de los
Muchachos of the Instituto de Astrof\'\i{}sica de Canarias.  The
author would like to thank the co-authors of Hynes et al.\
(\cite{Hynes:1999a}) and Hynes et al.\ (\cite{Hynes:2002a}) for their
assistance with the observations and many other valuable contributions
to these projects, and the referee, Marc Buie, for several helpful
suggestions.  This research has made use of the NASA Astrophysics Data
System Abstract Service.
\end{acknowledgements}

\appendix
\section{Analytic $\chi^2$ fitting of multiple profiles}
\label{App}
For the simple case considered here, the spatial profile of
blended sources is known {\it a priori} from a template source and it
is assumed that the offsets of these blended sources relative to the
template are known.  The extraction then reduces to finding the
flux scaling of each profile.  The optimal solution is equivalent to
minimising $\chi^2$, with care to reject any cosmic rays on the
profile.

Consider $N$ model profiles $P_{ij}$ ($j=1 \ldots N$) to be fitted to
an observed profile $D_i$ (variance $V_i$), yielding $N$
scalings $F_j$.  The badness of fit statistic is then:
\begin{equation}
\chi^2 = \sum_i \frac{(D_i - \sum_j F_j P_{ij})^2}{V_i}
\end{equation}
and the optimal solution requires $\partial\chi^2/\partial F_j= 0$ for
all $j$.  In general, these $N$ constraints will yield $N$
simultaneous equations with $N$ unknowns: $F_1 \ldots F_N$.  This can
be solved as a straightforward problem in linear algebra without
requiring a numerical minimisation.  Consider the two-profile
case.  For convenience write $F_1 \equiv f$, $F_2 \equiv g$, $P_{i,1}
\equiv p_i$ and $P_{i,2} \equiv q_i$.
%Then
%
%\begin{eqnarray}
%\chi^2 & = & \sum_i \frac{(D_i - fp_i -gq_i)^2}{V_i} \nonumber \\
%       & = & \sum_i \frac{D_i^2 + f^2P_i^2 + g^2q_i^2 - 
%			2fp_iD_i - 2gq_iD_i + 2fgp_iq_i}{V_i}
%\end{eqnarray}
%
%and so
%
%\begin{equation}
%\frac{\partial}{\partial f} \chi^2 = 0 \Rightarrow 
%\sum_i \frac{2fp_i^2 - 2p_iD_i +2gp_iq_i}{V_i} = 0
%\end{equation}
%
%\begin{equation}
%\frac{\partial}{\partial g} \chi^2 = 0 \Rightarrow 
%\sum_i \frac{2gq_i^2 - 2q_iD_i +2fp_iq_i}{V_i} = 0.
%\end{equation}
%
%These equations appear more complex than they are, since $f$ and $g$
%are independent of $i$ and are the only quantities that are not known
%or assumed {\it a priori}.  
Define the following statistics:
\begin{displaymath}
A = \sum_i p_i^2   / V_i \hspace{5mm}
B = \sum_i q_i^2   / V_i \hspace{5mm}
C = \sum_i p_i q_i / V_i
\end{displaymath}
\begin{equation}
D = \sum_i p_i D_i / V_i \hspace{5mm}
E = \sum_i q_i D_i / V_i.
\end{equation}
The two simultaneous equations then take the form
\begin{equation}
Af - D + Cg = 0 \hspace{5mm} Bg - E + Cf = 0
\label{SimEqnEqn}
\end{equation}
with solution
\begin{equation}
f = \frac{D/C - E/B}{A/C - C/B} \hspace{5mm} g = \frac{E/C - D/A}{B/C - C/A}.
\label{SolutionEqn}
\end{equation}

Formal errors can also be straightforwardly propagated from
Eqn.~\ref{SimEqnEqn} or \ref{SolutionEqn}:
\begin{displaymath}
\sigma_f^2 = \sum_i \left(\frac{\partial f}{\partial D_i}\right)^2V_i
= \frac{B}{AB-C^2}
\end{displaymath}
\begin{equation}
\sigma_g^2 = \sum_i \left(\frac{\partial g}{\partial D_i}\right)^2V_i
= \frac{A}{AB-C^2}
\end{equation}
\end{document}